\documentclass{ws-procs975x65}
\usepackage{graphics,graphicx,color}

\begin{document}

\title{QUANTUM HEISENBERG MODELS AND RANDOM LOOP REPRESENTATIONS}

\author{DANIEL UELTSCHI}

\address{Department of Mathematics,\\
University of Warwick,\\
Coventry, CV4 7AL, United Kingdom\\
E-mail: daniel@ueltschi.org}

\begin{abstract}
We review random loop representations for the spin-$\frac12$ quantum Heisenberg models, that are due to T\'oth (ferromagnet) and Aizenman--Nachtergaele (antiferromagnet). These representations can be extended to models that interpolate between the two Heisenberg models, such as the quantum {\it XY} model. We discuss the relations between long-range order of the quantum spins and the size of the loops. Finally, we describe conjectures about the joint distribution of the lengths of macroscopic loops, and of symmetry breaking.
\end{abstract}


\renewcommand{\thefootnote}{}
\footnote{\copyright{} 2012 by the author.}
\footnote{Supported in part by the EPSRC grant G056390/1.}

\bodymatter

\section{Introduction}

Many electronic properties of condensed matter systems can be described by quantum Heisenberg models. Electrons are assumed to be localized but their spins interact with those of neighboring electrons. The study of quantum spin systems is notoriously difficult, and although several important results have been obtained, much remains to be understood. In this article we review probabilistic representations. Probabilistic methods are not the most commonly employed in this subject, but they have proved useful nonetheless. A random walk representation has allowed Conlon and Solovej to obtain a lower bound on the free energy of the system.\cite{CS}
This result was then improved by T\'oth with the help of the random loop representation described below.\cite{Toth} Another loop representation, this time for the Heisenberg antiferromagnet, was proposed by Aizenman and Nachtergaele\cite{AN}. It allows to relate the 1D quantum model to 2D classical random cluster and Potts models. Recently, much progress has been achieved for another quantum model using similar probabilistic representations, the Ising model in transverse fields.\cite{Iof,CI,Gri,Bjo}

Random loop representations are also attractive {\it per se}. They allow to formulate open questions in a probabilistic setting, so that a whole new group of mathematicians can reflect upon them. One should hope that probabilistic methods can shed a light on several properties of quantum spin systems. The situation now is rather the opposite: Several results have been obtained for Heisenberg models (absence of spontaneous magnetization in one and two dimensions, and its occurrence in dimensions greater than two) using a genial combination of algebra and analysis, with insights from mathematical-physics. They provide stunning results when translated in the language of random loops.

The quantum Heinseberg models and their probabilistic representations are introduced in Section \ref{sec setting}. We describe in Section \ref{sec results} the theorem of Mermin and Wagner about the absence of spontaneous magnetization/macroscopic loops in dimensions one and two, and the theorem of Dyson, Lieb, and Simon about the occurrence of spontaneous magnetization in greater dimensions. Sections \ref{sec setting} and \ref{sec results} are mathematically rigorous. In the last Section \ref{sec heuristics} we describe the heuristics about the joint distribution of the lengths of macroscopic loops and we show that these conjectures are compatible with other conjectures concerning the breaking of U(1) or SO(3) symmetries. The claims of Section \ref{sec heuristics} are not proved.

\section{Quantum Heisenberg models and random loop representations}
\label{sec setting}

\subsection{Family of Heisenberg models}
Let $(\Lambda,{\mathcal E})$ be a finite graph (simple, no loops), where $\Lambda$ denotes the set of vertices, and ${\mathcal E}$ denotes the set of edges. We consider the Hilbert space ${\mathcal H}_{\Lambda} = \otimes_{x\in\Lambda} {\mathcal H}_{x}$, where each ${\mathcal H}_{x}$ is a copy of ${\mathbb C}^{2}$. Let $u \in [-1,1]$ be a parameter. We consider the following family of Hamiltonians:
\begin{equation}
H_{\Lambda}^{(u)} = -2 \sum_{\{x,y\} \in {\mathcal E}} \Bigl( S_{x}^{1} S_{y}^{1} + u S_{x}^{2} S_{y}^{2} + S_{x}^{3} S_{y}^{3} \Bigr).
\end{equation}
Here, $S_{x}^{i} = S^{i} \otimes {\mathrm{\texttt{Id}}}_{\Lambda \setminus \{x\}}$, where $S^{1}, S^{2}, S^{3}$ denote the usual spin operators for spin-$\frac12$ systems (Pauli matrices). The partition function of the model at inverse temperature $\beta$ is
\begin{equation}
\label{part fct}
Z^{(u)}(\beta,\Lambda) = {{\operatorname{Tr\,}}} {\rm e}^{-\beta H_{\Lambda}^{(u)}}.
\end{equation}
We use the usual notation $\langle \cdot \rangle$ to denote expectation of operators with respect to the Gibbs state, i.e.
\begin{equation}
\langle A \rangle = \frac1{Z^{(u)}(\beta,\Lambda)} {{\operatorname{Tr\,}}} A \, {\rm e}^{-\beta H^{(u)}_{\Lambda}}.
\end{equation}

This family of Hamiltonians contains several cases of interest.
\begin{itemize}
\item The case $u=1$ gives the {\bf Heisenberg ferromagnet}. In order to understand the physical motivation of this model, consider the Hilbert space for two spins at nearest-neighbors $x$ and $y$, and the symmetry group of spin rotations. This group yields the irreducible decomposition ${\mathcal H}_{x} \otimes {\mathcal H}_{y} = {\rm singlet} \oplus {\rm triplet}$. We want an interaction operator that is rotation invariant, so it must be of the form $c_{1} P_{\rm singlet} + c_{2} P_{\rm triplet}$. Up to constants and a shift by the identity operator, we get $\pm \vec S_{x} \cdot \vec S_{y}$, hence the two Heisenberg models.
\item Choosing $u=-1$, we get a model that is unitarily equivalent to the standard {\bf Heisenberg antiferromagnet} if the lattice is bipartite. The corresponding unitary operation consists of rotating all the spins of a sublattice by the angle $\pi$ around the second spin direction.
\item The case $u=0$ gives the {\bf {\it XY\/} model}. The standard representation involves interactions between spins in the axis 1 and 2, but the present choice is more suitable to the loop representation.
It is well-known that the {\it XY\/} model is equivalent to the model of hard-core bosons, see e.g.\ Refs~\refcite{KLS1,ALSSY}.
\end{itemize}

\subsection{Poisson point processes and loops}

\begin{figure}[h]
\begin{centering}
\begin{picture}(0,0)%
\epsfig{file=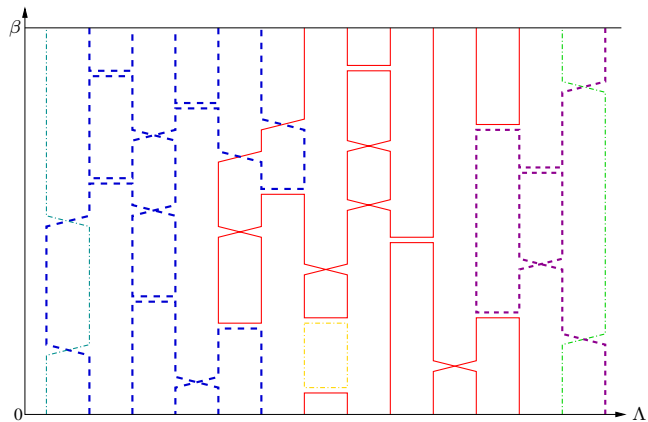}%
\end{picture}%
\setlength{\unitlength}{1776sp}%
\begingroup\makeatletter\ifx\SetFigFont\undefined%
\gdef\SetFigFont#1#2#3#4#5{%
  \reset@font\fontsize{#1}{#2pt}%
  \fontfamily{#3}\fontseries{#4}\fontshape{#5}%
  \selectfont}%
\fi\endgroup%
\begin{picture}(8700,5841)(976,-6490)
\put(9676,-6436){\makebox(0,0)[lb]{\smash{{\SetFigFont{7}{8.4}{\rmdefault}{\mddefault}{\updefault}{\color[rgb]{0,0,0}$\Lambda$}%
}}}}
\put(976,-1036){\makebox(0,0)[lb]{\smash{{\SetFigFont{7}{8.4}{\rmdefault}{\mddefault}{\updefault}{\color[rgb]{0,0,0}$\beta$}%
}}}}
\end{picture}%
\caption{A realization of the two Poisson processes on edges and their associated loop configuration $\omega$ on ${\mathcal E} \times [0,\beta]$ with periodic boundary conditions along the vertical direction. The number of loops, $|{\mathcal L}(\omega)|$, is 6 here. This picture is for $(\Lambda,{\mathcal E})$ being a line graph of 14 vertices, but arbitrary graphs can be considered.}
\label{fig loops}
\end{centering}
\end{figure}

The loop representation applies to general graphs $(\Lambda, {\mathcal E})$. We consider a Poisson point process on ${\mathcal E} \times [0,\beta]$ where
\begin{itemize}
\item crossings occur with intensity $\frac{1+u}2$.
\item bars occur with intensity $\frac{1-u}2$.
\end{itemize}
That is, the probability that a crossing occurs at edge $e \in {\mathcal E}$ and in the interval $[t,t+\varepsilon]$ is equal to $\frac{1+u}2 \varepsilon + O(\varepsilon^{2})$. Occurrences in disjoint intervals are independent events. Let $\rho_{\beta,\Lambda}^{(u)}$ denote the corresponding measure. Given a realization of this process, we define the loops in a natural way by following vertical lines, jumping to neighbors whenever a crossing or bars occurs (and continuing in the same direction in the case of a crossing, in the opposite direction in the case of bars). See Fig.\ \ref{fig loops} for an illustration.

Given a realization $\omega$ of the process $\rho_{\beta,\Lambda}^{(u)}$, let ${\mathcal L}(\omega)$ denote the set of loops, and $|{\mathcal L}(\omega)|$ the number of loops. The relevant probability measure is
\begin{equation}
\label{the measure}
\frac1{\tilde Z^{(u)}(\beta,\Lambda)} 2^{|{\mathcal L}(\omega)|} {{\rm d}}\rho_{\beta,\Lambda}^{(u)}(\omega)
\end{equation}
where the normalization is
\begin{equation}
\tilde Z^{(u)}(\beta,\Lambda) = \int 2^{|{\mathcal L}(\omega)|} {{\rm d}}\rho_{\beta,\Lambda}^{(u)}(\omega).
\end{equation}

The case $u=1$ and without the term $2^{|{\mathcal L}(\omega)|}$ is often called the random interchange model. The size of its loops has been the object of several recent studies when the graph is a tree \cite{Ang,Ham1,Ham2} and the complete graph \cite{Sch,Ber,AK,BK}.

\subsection{Relations between Heisenberg models and random loops}
A major advantage of the loop model is that it represents the Heisenberg models faithfully. Not only 
are the partition functions identical, but the spin correlations are given by the natural correlation functions in the loop model, namely that sites belong to the same loop. The following theorem has been progressively proved in Refs~\refcite{Toth,AN,Uel}.

\begin{theorem}
\begin{itemize}
\item[(a)] The partition functions of quantum spins and random loops are identical:
\[
Z^{(u)}(\beta,\Lambda) = \tilde Z^{(u)}(\beta,\Lambda).
\]
\item[(b)] Spin correlations in directions 1 and 3 are given by
\[
\langle S_{x}^{1} S_{y}^{1} \rangle = \langle S_{x}^{3} S_{y}^{3} \rangle = \frac14 {\mathbb P}^{(u)}_{\beta,\Lambda} \bigl( (x,0) \; {\rm and} \; (y,0) {\rm \; belong \; to \; the \; same \; loop} \bigr).
\]
\item[(c)] Spin correlations in the direction 2 also have a loop counterpart:
\[
\langle S_{x}^{2} S_{y}^{2} \rangle = \frac14 {\mathbb P}_{\beta,\Lambda}^{(u)}(E_{xy}^{+}) - \frac14 {\mathbb P}_{\beta,\Lambda}^{(u)}(E_{xy}^{-}).
\]
\end{itemize}
\end{theorem}

Here, $E_{xy}^{+}$ denotes the event where $(x,0)$ and $(y,0)$ belong to the same loop, and the loop is moving in the same vertical direction at those points. $E_{xy}^{-}$ is the similar event where the loop is moving in opposite vertical direction. Notice that $E_{xy}^{+} \cup E_{xy}^{-}$ is the event where $(x,0)$ and $(y,0)$ belong to the same loop. An immediate consequence of the theorem is that
\begin{equation}
\label{comp spin directions}
\bigl| \langle S_{x}^{2} S_{y}^{2} \rangle \bigr| \leq \langle S_{x}^{1} S_{y}^{1} \rangle = \langle S_{x}^{3} S_{y}^{3} \rangle,
\end{equation}
with equality if and only if $u=\pm1$.

Macroscopic loops are related to two physical properties of the system, namely spontaneous magnetization and magnetic susceptibility. This is stated in the following theorem. A more precise result, and a proof, can be found in Ref.~\refcite{Uel}. Let $L_{(0,0)}$ denote the length of the loop that contains the point $(0,0) \in \Lambda \times [0,\beta]$.

\newpage
\begin{theorem}
\begin{itemize}
\item[(a)] Relation between macroscopic loops and magnetic susceptibility:
\[
{\mathbb E}^{(u)}_{\beta,\Lambda} \Bigl( \frac{L_{(0,0)}}{\beta |\Lambda|} \Bigr) = \frac4{|\Lambda|^{2}} \frac{\partial^{2}}{\partial \eta^{2}} \log {{\operatorname{Tr\,}}} \, {\rm e}^{-\beta H_{\Lambda}^{(u)} + \eta \sum_{x\in\Lambda} S_{x}^{3}} \, \Big|_{\eta=0}.
\]
\item[(b)] Relation between macroscopic loops and spin correlation functions:
\[
\frac4{|\Lambda|^{2}} \sum_{x,y\in\Lambda} \langle S_{x}^{3} S_{y}^{3} \rangle - \sqrt{ \frac{2d(1-u)}{|\Lambda|}} \leq {\mathbb E}^{(u)}_{\beta,\Lambda} \Bigl( \frac{L_{(0,0)}}{\beta|\Lambda|} \Bigr) \leq \frac4{|\Lambda|^{2}} \sum_{x,y\in\Lambda} \langle S_{x}^{3} S_{y}^{3} \rangle.
\]
\end{itemize}
\end{theorem}

\section{Results about spontaneous magnetization/macroscopic loops}
\label{sec results}

There are two major results, namely the absence of spontaneous magnetization in dimensions 1 and 2, and its occurrence in dimensions 3 and more, if the temperature is low enough.

The first result is originally due to Mermin and Wagner,\cite{MW} and to Fisher and Jasnow for the decay of correlations.\cite{FJ} Their methods used Bogolubov's inequality and the translation invariance in ${\mathbb Z}^{1}$ or ${\mathbb Z}^{2}$. But the latter property is not essential. The theorem presented here and its proof are inspired by Fr\"ohlich and Pfister \cite{FP}, and Nachtergaele \cite{Nac}. It is actually less general than those of Refs~\refcite{FP,Nac} but its proof is simpler and it avoids the theory of abstract KMS states.

Given a connected graph $(\Lambda,{\mathcal E})$, we let $d(x,y)$ denote the length of the minimal connected path from $x$ to $y$.

\begin{theorem}
\label{thm MW}
Assume that there exists a constant $C$ such that for any $x\in\Lambda$ and any integer $k$,
\[
\#\bigl\{ y : d(x,y) = k \bigr\} \leq Ck.
\]
Then there exists a constant $K$, that depends on $C$ and $\beta$ but that does not depend on $(\Lambda,{\mathcal E})$, such that for all $x,y \in \Lambda$,
\[
0 \leq \langle S_{x}^{3} S_{y}^{3} \rangle \leq \frac{K}{\sqrt{\log d(x,y)}}.
\]
\end{theorem}

Notice that the bound also applies to correlations in other spin directions, using Eq.\ \eqref{comp spin directions}. It is not hard to check that the theorem rules out the possibility of spontaneous magnetization on ${\mathbb Z}^{2}$ and in other two-dimensional graphs, at arbitrary positive temperatures.

\begin{proof}[Sketch proof]
Let $n = d(x,y)$. We consider the following unitary operation that rotates the spin at $x$ but not at $y$, and that interpolates smoothly between them:
\begin{equation}
U = \prod_{z\in\Lambda} \, {\rm e}^{{{\rm i}} \phi_{z} S_{z}^{2}},
\end{equation}
with
\begin{equation}
\phi_{z} = \begin{cases} \bigl( 1 - \frac{\log(d(x,z)+1)}{\log(n+1)} \bigr) \pi & \text{if } d(x,z) \leq n, \\ 0 & \text{otherwise.} \end{cases}
\end{equation}
Since $U^{*} S_{x}^{3} U = -S_{x}^{3}$ and $U^{*} S_{y}^{3} U = S_{y}^{3}$, we have
\begin{equation}
\label{voila}
{{\operatorname{Tr\,}}} S_{x}^{3} S_{y}^{3} \, {\rm e}^{-\beta H_{\Lambda}^{(u)}} = - {{\operatorname{Tr\,}}} S_{x}^{3} S_{y}^{3} \, {\rm e}^{-\beta U^{*} H_{\Lambda}^{(u)} U}.
\end{equation}
If we could remove the $U$'s in the right side, the correlation would be zero. We show that it can indeed be done, at least approximately. We have
\begin{equation}
\begin{split}
U^{*} H_{\Lambda}^{(u)} U = &-\sum_{\{z,z'\} \in {\mathcal E}} U^{*} \bigl( S_{z}^{1} S_{z'}^{1} + u S_{z}^{2} S_{z'}^{2} + S_{z}^{3} S_{z'}^{3} \bigr) U \\
= &-\sum_{\{z,z'\} \in {\mathcal E}} \, {\rm e}^{-{{\rm i}} \phi_{z} (S_{z}^{2} + S_{z'}^{2})} \, \Bigl( S_{z}^{1} \, {\rm e}^{-{{\rm i}} (\phi_{z'}-\phi_{z}) S_{z'}^{2}} \, S_{z'}^{1} \, {\rm e}^{{{\rm i}} (\phi_{z'}-\phi_{z}) S_{z'}^{2}} \\
& + u S_{z}^{2} S_{z'}^{2} + S_{z}^{3} \, {\rm e}^{-{{\rm i}} (\phi_{z'}-\phi_{z}) S_{z'}^{2}} \, S_{z'}^{3} \, {\rm e}^{{{\rm i}} (\phi_{z'}-\phi_{z}) S_{z'}^{2}} \, \Bigr) \, {\rm e}^{-{{\rm i}} \phi_{z} (S_{z}^{2} + S_{z'}^{2})}.
\end{split}
\end{equation}
Observe that ${\rm e}^{-{{\rm i}}\alpha S_{z'}^{2}} \, S_{z'}^{1} \, {\rm e}^{{{\rm i}}\alpha S_{z'}^{2}} = S_{z'}^{1} - \alpha S_{z'}^{3} + O(\alpha^{2})$, and a similar identity for the rotation of $S_{z'}^{3}$. Using the invariance of $S_{z}^{1} S_{z'}^{1} + S_{z}^{3} S_{z'}^{3}$ and $S_{z}^{1} S_{z'}^{3} - S_{z}^{3} S_{z'}^{1}$ under rotations around the second direction, we get
\begin{equation}
\label{rotated Ham}
U^{*} H_{\Lambda}^{(u)} U = H_{\Lambda}^{(u)} + \sum_{\{z,z'\} \in {\mathcal E}} (\phi_{z'}-\phi_{z}) (S_{z}^{1} S_{z'}^{3} - S_{z}^{3} S_{z'}^{1}) + \sum_{\{z,z'\} \in {\mathcal E}} O( (\phi_{z'}-\phi_{z})^{2} ).
\end{equation}
We now use the identity ${{\operatorname{Tr\,}}} A \, {\rm e}^{-H} - {{\operatorname{Tr\,}}} \, {\rm e}^{-H} = \lim_{s\to0} \frac1s {{\operatorname{Tr\,}}} \, {\rm e}^{-H+sA}$ and Klein's inequality, so as to get
\begin{equation}
\label{la borne}
\begin{split}
\frac1{Z^{(u)}(\beta,\Lambda)} &{{\operatorname{Tr\,}}} S_{x}^{3} S_{y}^{3} \Bigl( \, {\rm e}^{-\beta H_{\Lambda}^{(u)}} - {\rm e}^{-\beta U^{*} H_{\Lambda}^{(u)} U} \, \Bigr) =\\
&= \frac1{s \, Z^{(u)}(\beta,\Lambda)} {{\operatorname{Tr\,}}} \Bigl( \, {\rm e}^{-\beta H_{\Lambda}^{(u)} + s S_{x}^{3} S_{y}^{3}} - {\rm e}^{-\beta U^{*} H_{\Lambda}^{(u)} U + s S_{x}^{3} S_{y}^{3}} \, \Bigr) + O(s)\\
&\leq \frac1{s \, Z^{(u)}(\beta,\Lambda)} {{\operatorname{Tr\,}}} \bigl( U^{*} H_{\Lambda}^{(u)} U - H_{\Lambda}^{(u)} \bigr) \, {\rm e}^{-\beta H_{\Lambda}^{(u)} + s S_{x}^{3} S_{y}^{3}} + O(s).
\end{split}
\end{equation}
We use Eq.\ \eqref{rotated Ham} for the difference inside the trace. The middle term in Eq.\ \eqref{rotated Ham} gives 0 because of the symmetry $\prod_{z\in\Lambda} \, {\rm e}^{{{\rm i}} \pi S_{z}^{3}}$, that sends $S_{z}^{1}$ onto $-S_{z}^{1}$. We have
\begin{equation}
|\phi_{z'}-\phi_{z}| \leq \frac{\text{const}}{d(x,z) \log(n+1)}
\end{equation}
for $d(x,z) \leq n$, 0 otherwise. Using the assumption of the theorem, the higher order correction in \eqref{rotated Ham} is bounded by
\begin{equation}
\label{laborne}
\text{const} \sum_{k=1}^{n} k \frac1{(k \log(n+1))^{2}} \leq \frac{\text{const}}{\log(n+1)}.
\end{equation}
Combining Eqs \eqref{voila}, \eqref{la borne}, and \eqref{laborne}, we get for any $s \in (0,1]$
\begin{equation}
\begin{split}
0 \leq \langle S_{x}^{3} S_{y}^{3} \rangle &= \frac1{2 Z^{(u)}(\beta,\Lambda)} {{\operatorname{Tr\,}}} S_{x}^{3} S_{y}^{3} \Bigl( \, {\rm e}^{-\beta H_{\Lambda}^{(u)}} - {\rm e}^{-\beta U^{*} H_{\Lambda}^{(u)} U} \, \Bigr) \\
&\leq \frac{\text{const}}{s \log(n+1)} + O(s).
\end{split}
\end{equation}
We choose $s = \frac1{\sqrt{\log(n+1)}}$ in Eq.\ \eqref{la borne} and we get the claim of the theorem.
\end{proof}

The second theorem is a positive result about the occurrence of spontaneous magnetization and it was proposed by Dyson, Lieb, and Simon \cite{DLS}. It was a ``breakthrough based on another breakthrough'' as Nachtergaele wrote it \cite{Nac1}. It extended to quantum systems the method of infrared bounds developed by Fr\"ohlich, Simon, and Spencer for the classical Heisenberg model.\cite{FSS} The result was initially proved for spin-$\frac12$ systems for $d\geq5$, but it was extended to $d\geq3$ by Kennedy, Lieb, and Shastry \cite{KLS1}, following observations by Neves and Perez \cite{NP}. Notice that it applies only to the antiferromagnetic case $u \in [-1,0]$. It is a notoriously unsolved problem to extend this result to ferromagnetic systems. The theorem is elegantly formulated in terms of macroscopic loops.

\begin{theorem}
\label{thm DLS}
Let $(\Lambda,{\mathcal E})$ be a cubic box in ${\mathbb Z}^{d}$ with $d\geq3$, even side lengths, and periodic boundary conditions. Let $u \in [-1,0]$. Then there exist $\beta_{0}<\infty$ and $\eta>0$, independent of the size of the box, such that
\[
{\mathbb E}_{\beta,\Lambda}^{(u)} \Bigl( \frac{L_{(0,0)}}{\beta |\Lambda|} \Bigr) \geq \eta.
\]
\end{theorem}

It is possible to prove this theorem directly in the random loop setting without referring to the quantum framework.\cite{Uel}

\section{Heuristics and conjectures}
\label{sec heuristics}

\subsection{Joint distribution of macroscopic loops}

This section is based on the discussion of Ref.~\refcite{GUW}.
Recall that a partition of the interval $[0,1]$ is a sequence $(\lambda_{1},\lambda_{2},\lambda_{3},\dots)$ of nonnegative, decreasing numbers such that $\sum_{i} \lambda_{i} = 1$. Let $L_{1}(\omega), L_{2}(\omega)$ denote the lengths of the loops of $\omega$ in decreasing order. Then $\bigl(\frac{L_{1}(\omega)}{\beta |\Lambda|}, \frac{L_{2}(\omega)}{\beta |\Lambda|}, \dots\bigr)$ is a random partition of $[0,1]$. As $|\Lambda|\to\infty$, we know that $\frac{L_{1}(\omega)}{\beta |\Lambda|}$ does not go to 0 if $d\geq3$ and $\beta$ is large enough (see Theorem \ref{thm DLS}). On the other hand, we should expect that a fraction of the domain $\Lambda \times [0,\beta]$ belongs to loops of length $\beta$, and more generally, to loops of bounded lengths. The first conjecture is a strong law of large numbers, that states that only finite and macroscopic loops are present in the system.

\begin{conjecture}
\label{conj one}
There exists $\nu \in [0,1]$ such that
\begin{align}
&\lim_{K\to\infty} \lim_{|\Lambda|\to\infty} \sum_{i : L_{i}(\omega) < K} \frac{L_{i}(\omega)}{\beta |\Lambda|} = 1 - \nu. & \text{(finite loops)} \nonumber\\
&\lim_{k\to\infty} \lim_{|\Lambda|\to\infty} \sum_{i =1}^{k} \frac{L_{i}(\omega)}{\beta |\Lambda|} = \nu. & \text{(macroscopic loops)} \nonumber
\end{align}
\end{conjecture}

We know that $\nu=0$ in $d=1,2$, and at high temperature in $d\geq3$. But we should have $\nu>0$ if $d\geq3$ and if the temperature is low enough. The conjecture also implies that $\bigl(\frac{L_{1}(\omega)}{\beta |\Lambda|}, \frac{L_{2}(\omega)}{\beta |\Lambda|}, \dots\bigr)$ converges in distribution to a random partition of $[0,\nu]$.

In order to formulate a conjecture for the limiting distribution, we need to recall the definition of the Poisson-Dirichlet (PD) distribution. This is best done with the help of the closely related Griffiths-Engen-McCloskey (GEM) distribution. Let $X_{1}, X_{2}, \dots$ be i.i.d.\ random variables of law beta$(\theta)$ (that is, $X_{i}$ takes values in $[0,1]$ and ${\mathbb P}(X_{i}>s) = (1-s)^{\theta}$ for $0<s<1$). Then the random sequence
\begin{equation}
\label{random seq}
\Bigl( X_{1}, \; (1-X_{1}) X_{2}, \; (1-X_{1}) (1-X_{2}) X_{3}, \; \dots \Bigr)
\end{equation}
is distributed according to GEM$(\theta)$. This is the ``stick breaking'' construction, since $(1-X_{1}) \dots (1-X_{k})$ is what is left of the interval after chopping off $k$ pieces. Rearranging these numbers in decreasing order, we get a random partition with distribution PD$(\theta)$.

\begin{conjecture}
\begin{itemize}
\item[(a)] If $u = \pm1$, $\bigl(\frac{L_{1}(\omega)}{\beta |\Lambda| \nu}, \frac{L_{2}(\omega)}{\beta |\Lambda| \nu}, \dots\bigr)$ converges in distribution to Poisson-Dirichlet(2).
\item[(b)] If $-1<u<1$, $\bigl(\frac{L_{1}(\omega)}{\beta |\Lambda| \nu}, \frac{L_{2}(\omega)}{\beta |\Lambda| \nu}, \dots\bigr)$ converges in distribution to Poisson-Dirichlet(1).
\end{itemize}
\end{conjecture}

The mechanism behind the Poisson-Dirichlet distributions of the lengths of macroscopic loops is indirect but very general. The explanation is motivated by Schramm's work on the composition of random transpositions in the complete graph \cite{Sch}, proving a conjecture of Aldous. This is explained in details in Ref.~\refcite{GUW} in the case $u=\pm1$, and it proceeds as follows.
\begin{itemize}
\item Introduce a stochastic process such that the equilibrium measure $2^{|{\mathcal L}(\omega)|} {{\rm d}}\rho_{\beta,\Lambda}^{(u)}(\omega)$ is the invariant measure.
\item This yields an effective split-merge process on partitions.
\item The invariant measure of the split-merge process is Poisson-Dirichlet \cite{MZZ}.
\end{itemize}
The Markov process is quite natural and is defined as follows.
\begin{itemize}
\item A new edge-time $(e,t)$ appears at rate $\sqrt2 {{\rm d}} t$ if its appearance causes a loop to split, and at rate $1/\sqrt2 {{\rm d}} t$ if it causes two loops to merge.
\item An edge-time already present disappears at rate $\sqrt2$ if its removal causes a loop to split, and at rate $1/\sqrt2$ if it causes two loops to merge.
\end{itemize}

By considering all possible cases, we can check the {\bf detailed balance condition}:
\begin{equation}
\rho({{\rm d}}\omega) 2^{|{\mathcal L}(\omega)|} p(\omega,{{\rm d}}\omega') = \rho({{\rm d}}\omega') 2^{|{\mathcal L}(\omega')|} p(\omega',{{\rm d}}\omega).
\end{equation}
and since the process is ergodic, the measure $2^{|{\mathcal L}(\omega)|} \rho_{\beta,\Lambda}^{(u)}({{\rm d}}\omega)$ is the unique invariant measure (up to a normalization).

There is an important distinction between the case $u=\pm1$ on the one hand, and the case $-1<u<1$ on the other hand. In the first case, any local change results in a merge or in a split (see Ref.~\refcite{GUW} for details). But in the case $-1<u<1$, where both crossings and bars are present, it is possible that local changes do not break the loop, but only modify its internal order (think of $0 \leftrightarrow 8$) The probability of splitting loops is then halved, and the effective split-merge process has stationary distribution PD(1) instead of PD(2).

Models of spatial permutations are closely related. The occurrence of the Poisson-Dirichlet distribution can be proved in the ``annealed'' model where positions are averaged upon \cite{BU}. The lattice model is harder to study rigorously, but the mechanisms described above have been verified numerically \cite{GLU}.

\subsection{Macroscopic loops vs symmetry breaking}

Heisenberg models have natural rotation symmetries, namely SO(3) in the case $u = \pm1$ and U(1) in the case $-1<u<1$. They are expected to be broken in $d\geq3$ and at temperatures low enough. These symmetries are not apparent in the loop representations. But we show here that there is a good reason why two different Poisson-Dirichlet distributions appear, PD(2) and PD(1), in these different situations.

We consider the two-point correlation function $\langle S_{x}^{3} S_{y}^{3} \rangle$ with $x,y$ far apart. If $u = +1$, we expect that
\begin{equation}
3 \langle S_{x}^{3} S_{y}^{3} \rangle = \langle \vec S_{x} \cdot \vec S_{y} \rangle = \frac1{4\pi} \int_{{\mathbb S}^{2}} \langle \vec S_{x} \cdot \vec S_{y} \rangle_{\vec\Omega} {{\rm d}}\vec\Omega,
\end{equation}
where $\langle \cdot \rangle_{\vec\Omega}$ is the pure state obtained by adding the external magnetic field $\varepsilon  \vec\Omega \cdot\sum_{x} \vec S_{x}$ and by letting $\varepsilon \searrow 0$ after taking the thermodynamic limit. Using rotation invariance, it is enough to consider $\vec\Omega = \vec e_{3}$. Then $\langle S_{x}^{1} S_{y}^{1} \rangle_{\vec e_{3}} = \langle S_{x}^{2} S_{y}^{2} \rangle_{\vec e_{3}} = 0$, and, as $\|x-y\|\to\infty$, $\langle S_{x}^{3} S_{y}^{3} \rangle_{\vec e_{3}}= \langle S_{x}^{3} \rangle_{\vec e_{3}} \langle S_{y}^{3} \rangle_{\vec e_{3}} = \frac{\nu^{2}}4$, with $\nu$ the number appearing in Conjecture \ref{conj one}. Then $\langle S_{x}^{3} S_{y}^{3} \rangle = \frac{\nu^{2}}{12}$.

The case $u=-1$ on bipartite lattices is similar. But one should perform the symmetry operation that gives the antiferromagnet $+\sum \vec S_{x} \cdot \vec S_{y}$ in order to use rotation invariance. We also get $\langle S_{x}^{3} S_{y}^{3} \rangle = \frac{\nu^{2}}{12}$.

If $-1<u<1$, we rather expect that
\begin{equation}
2 \langle S_{x}^{3} S_{y}^{3} \rangle = \langle S_{x}^{1} S_{y}^{1} +S_{x}^{3} S_{y}^{3} \rangle = \frac1{2\pi} \int_{{\mathbb S}^{1}} \langle S_{x}^{1} S_{y}^{1} +S_{x}^{3} S_{y}^{3} \rangle_{\vec\Omega} {{\rm d}}\vec\Omega,
\end{equation}
where $\vec\Omega$ is in the plane $(e_{1},e_{3})$. It is again enough to consider $\vec\Omega = \vec e_{3}$, and $\langle S_{x}^{1} S_{y}^{1} \rangle_{\vec e_{3}} = \langle S_{x}^{2} S_{y}^{2} \rangle_{\vec e_{3}} = 0$.  Since $\langle S_{x}^{3} S_{y}^{3} \rangle_{\vec e_{3}}= \langle S_{x}^{3} \rangle_{\vec e_{3}} \langle S_{y}^{3} \rangle_{\vec e_{3}} = \frac{\nu^{2}}4$ as $\|x-y\|\to\infty$, we find $\langle S_{x}^{3} S_{y}^{3} \rangle = \frac{\nu^{2}}{8}$.

Let us now calculate these correlations in the random loop representation. For $x,y$ very far apart, the probability that they belong to the same loop is the same as the probability that two points in $[0,1]$ belong to the same element of the partition, multiplied by $\nu^{2}$. This probability is easy to compute using the GEM distribution. Recall that if $X$ is a beta$(\theta)$ random variable, the expectation of $X^{2}$ is equal to $\frac{2}{(\theta+1) (\theta+2)}$, and the expectation of $(1-X)^{2}$ is equal to $\frac\theta{\theta+2}$. Using the definition \eqref{random seq} for the GEM random partition, and summing over the probabilities that the numbers both belong to the $k$th element, we find
\begin{equation}
\begin{split}
{\mathbb P}( s,t \text{ belong to the same element}) &= \sum_{k\geq1} {\mathbb E} \bigl( (1-X_{1})^{2} \dots (1-X_{k-1})^{2} X_{k}^{2} \bigr) \\
&= \sum_{k\geq1} \Bigl( \frac\theta{\theta+2} \Bigr)^{k-1} \frac2{(\theta+1) (\theta+2)} \\
&= \frac1{\theta+1}.
\end{split}
\end{equation}
This gives
\begin{equation}
{\mathbb P}^{(u)}_{\beta,\Lambda} \bigl( \text{$(x,0)$ and $(y,0)$ belong to the same loop} \bigr) = \begin{cases} \nu^{2}/3 & \text{if } u = \pm1, \\ \nu^{2}/2 & \text{if } -1<u<1. \end{cases}
\end{equation}
This is indeed equal to $4 \langle S_{x}^{3} S_{y}^{3} \rangle$, as found above.

These heuristics calculations show that the conjectures about the Poisson-Dirichlet distributions of the lengths of the loops are compatible with the conjectures about the breaking of rotation invariance SO(3) or U(1).

\end{document}